\documentclass[10pt,conference,letterpaper]{IEEEtran}

\usepackage{times}
\usepackage[]{todonotes}
\usepackage[cmex10]{amsmath}
\usepackage{graphicx}
\usepackage{array}
\usepackage{mdwtab}
\usepackage{hyperref}
\usepackage{booktabs}
\usepackage{balance}
\usepackage{flushend}
\usepackage{bbding}

\usepackage{makecell}

\begin{document}
\pagenumbering{arabic} 
\pagestyle{plain}



\title{The Benefits of Deploying Smart Contracts on Trusted Third Parties}

\author{
\IEEEauthorblockN{Carlos Molina--Jimenez}
\IEEEauthorblockA{Computer Laboratory\\
University of Cambridge, UK\\
carlos.molina@cl.cam.ac.uk}\\   

\and

\IEEEauthorblockN{Ioannis Sfyrakis}
\IEEEauthorblockA{School of Computing\\
Newcastle University, UK\\
ioannis.sfyrakis@ncl.ac.uk}\\

\and

\IEEEauthorblockN{Linmao Song}
\IEEEauthorblockA{Independent researcher\\
linmao.song@gmail.com}

\and

\IEEEauthorblockN{Hazem Danny Al Nakib}
\IEEEauthorblockA{Independent fintech and regtech expert\\
hazemdanny.nakib@gmail.com}

\and

\IEEEauthorblockN{\hspace{6cm} Jon Crowcroft}
\IEEEauthorblockA{\hspace{6cm}Computer Laboratory\\
\hspace{6cm}University of Cambridge, UK\\
\hspace{6cm}jon.crowcroft@cl.cam.ac.uk\\}

}

%
%
%
%
%

\maketitle
\thispagestyle{empty}
%
%

\begin{abstract}
The hype about Bitcoin has overrated the potential of smart contracts 
deployed on--blockchains (on--chains) and underrated the potential of smart contracts 
deployed on--Trusted Third Parties (on--TTPs). As a result, current research 
and development in this field is focused mainly on smart contract applications 
that use on-chain smart contracts. We argue that there 
is a large class of smart contract applications where on--TTP smart 
contracts are a better alternative. The problem with on--chain smart contracts 
is that the fully decentralised model and indelible append--only data model 
followed by blockchains introduces several engineering problems that are 
hard to solve. In these situations, the inclusion of a TTP (assuming that 
the application can tolerate its inconveniences) instead of a blockchain to 
host the smart contract simplifies the problems and offers pragmatic solutions.
The intention and contribution of this paper is to shed some light on this issue. 
We use a hypothetical use case of a car insurance application to illustrate 
technical problems that are easier to solve with on--TTP smart contracts than 
with on--chain smart contracts\footnote{This document is a copy of the
one uploaded to Research Gate in April 2019, except for typo
corrections and a few lines added to the end of Sections D  and VI.}.  

\begin{IEEEkeywords}
smart contracts, on--blockchain, off--blockchain, contract compliance, 
contract enforcement, contractual rights and obligations.
\end{IEEEkeywords}


\end{abstract}

\section{Introduction} 
\label{introduction}
The publication of the Bitcoin paper by~\cite{Satoshi2008} in 2008 has motivated the exploration of blockchains and smart contracts in the development of innovative applications. There is a global consensus on the potential of smart contracts for building innovative applications across various  sectors of industry and government~\cite{Hileman2017}. We agree with this view and consider that there are smart contract applications that are conveniently implemented on the basis of smart contracts deployed on--chain like those on the Ethereum~\cite{Ethereum2017}, 
Hyperledger Fabric~\cite{HyperledgerHome}  or Bitcoin~\cite{BitcoinHome} blockchains. However, we feel that the potential of on--chain smart contracts has been overrated. There are applications where the use of on-blockchain smart contracts is  overkill and others  where  they  have  a  detrimental effect to the extent of rendering the application unpractical or unimplementable~\cite{Morgen2018,Scriber2018,Valentina2018}. In these situations, developers should consider the use of smart contracts deployed on Trusted Third Parties. It seems that the confusion is caused by the novelty of the technology and lack of global consensus about what smart contracts and blockchains are. Before continuing the discussion we will present the definitions that will be followed in subsequent sections of this paper.

\subsection{Blockchains}
\label{blockchain}
To software developers, a blockchain is a piece of middle- ware that includes built--in services that can be used for the development of distributed applications without the involvement of a component (for example, an authentication or a database server) acting as a central control. The salient services that blockchains can offer are indelible append--only storage, public--key--based access control to the execution of 
operations, public auditability of records and, notably, consensus services of some discipline  (normally,  eventual  consensus).  A great variety of applications can be built on the basis of blockchains; smart contracts are the best known but not the only one. At an implementation level, a blockchain can be defined as a distributed data structure (known also as the ledger) composed of an ordered, back--linked list (chain) of blocks. Blocks are added to the list following an append--only model, and only when a set of nodes  that hold local replicas of the current state of the blockchain reach consensus about its next global state. The blocks are normally used for storing records of transactions executed by parties that do not necessarily trust each other and underpin the peer--to--peer execution model of the transactions. The transactions are operations aimed at altering the current state of the blockchain and are  not necessarily bank transactions.
The building and maintenance  of  a  blockchain  requires  the combination of several technologies, consequently, different conceptual models have been suggested for modelling blockchains. For instance, in~\cite{KenjiSaito2016} a blockchain is regarded as   a probabilistic state machine to emphasise the probabilistic nature of the consensus algorithms that are  used in blockchains.

\subsection{Smart contracts}
\label{smartcontract}
Intuitively, a smart contract is aimed at being the digital version of a conventional 
commercial contract that is written in a human language, for example,  English. 
By  digital  we  mean that the  smart  contract  is  a  piece  of  computer  code  
that  can  be executed to monitor or enforce the  clauses  stipulated  in the 
conventional contract, programmatically at run--time. For example, a financial 
contract for a loan agreement  between a borrower and 
lender might stipulate that at  the  request  of  the  borrower,  the lender will advance GBP 100.00 to the borrower  within  five days of receiving the request. Being an executable program, a smart contract is expected to be deployed on  infrastructure that can provide it with a run--time environment, including, library support, facilities for managing the smart contract (initiate, stop, terminate, update, etc.) and databases for storing records about the inputs sent to the smart contracts and outputs produced by it. The main responsibilities of the smart contract are i) to act as a contract monitor that observes the interaction between the contracting parties; ii) to determine whether each operation (for example, request loan, honour loan request, decline loan request, etc.) that the contracting parties execute  is  compliant  or non-compliant with the terms stipulated in the contract and in accordance with the rights and obligations inherent in the contract and iii) to produce a verdict. Notice that in some existing smart contract applications, the declaration of the verdict is directly and intricately associated with an action (e.g., collect payment) that   is executed when the verdict is contract compliant. However, we believe that for the sake of modularity it makes sense to separate the two acts and focus on the verdict---the most fundamental operation that smart contracts need to perform. We regard the action as an arbitrary reaction that can be immediately or eventually executed by the smart  contract  itself or by a different component implemented in a different software layer. In many applications, the action consists of allowing the execution of  the  contractual  operation  when  it is declared contract compliant or preventing its execution  when the operation is declared non--contract compliant by the smart contract. In this approach, the smart contract acts as a contract enforcer. The question that arises here is where to deploy the smart contract.

\subsection{On--blockchain smart contracts}
\label{smartcontractdeployment}
Smart contracts can be deployed on--chain. The term smart contract was coined by 
Szabo in 1997~\cite{Szabo1997} but it did not attract commercial interest until 2008  with the 
commercial success of Bitcoin as a cryptocurrency platform~\cite{Satoshi2008}. In Bitcoin~\cite{AndreasAntonopoulos2017}, 
smart contracts are implemented in an opcode stack--based script non--Turing complete 
language~\cite{BitcoinScript} and deployed on the Bitcoin blockchain. They were originally conceived 
for mediating the execution of payment operations in BTCs (Bitcoin's cryptocurrency). 
Ethereum~\cite{AndreasAntonopoulos2018}, currently a leading blockchain platform, extends the idea of Bitcoin's smart contracts and implements a blockchain that supports smart contracts written in  
Turing--complete  languages,  such as Solidity~\cite{Solidity2017}. As demonstrated by numerous Bitcoin and Ethereum use cases, in several situations, smart contracts can be conveniently deployed on--chain, but it is crucial to appreciate that blockchains and smart contracts are two independent technologies. The deployment of smart contracts on--blockchain is only one out of several deployment alternatives. To appreciate the point, it is worth taking into account that smart contracts predate blockchains by several years. Research on smart contracts was pioneered by Minsky  in  the  mid--80s~\cite{Minsky1985} and followed by Marshall~\cite{Lindsay1993}. In early works, smart contracts were referred to as electronic contracts, executable contracts, digital contracts and other similar names. Since the usage of the term smart contract  is so widespread now,  we  will follow it in the rest of this document.

\subsection{On--Trusted Third Parties smart contracts}
\label{onTTPdeployment}
Smart contracts can be deployed on--Trusted Third Parties (on--TTP). 
We  define a TTP as an institution that in addition   to having the 
technical infrastructure for hosting smart contracts, has earned trust, 
reputation and authority. E--commerce applications routinely make use 
of TTP services, examples include payment gateways, certification 
authorities, time--stamping services, custodian services, settlement 
services and various brokerage services~\cite{Mainelli2015}.
In  smart  contract  applications, a  TTP  is logically located between the 
two business partners and is trusted to host and run a single instance of the smart 
contract. The smart  contract observes all the operations that the partners 
execute and keeps undeniable records about them. Pioneering work 
on smart contracts~\cite{Minsky1985,Lindsay1993} was mainly focused 
on on--TTP deployment.  

A strong assumption with a TTP--based deployment is that
the TTP is trusted not to alter the code of the contract 
or abuse the sensitive data that the contractual parties expose 
to it. This issue can be addressed with the assistance of recently
emerging trusted hardware technologies such as Intel SGX~\cite{VictorCostan2016}
and ARM TrustZone~\cite{Sandro2019}. With this approach, the
contract code can be deployed within the trusted environments
that trusted hardware offer (called enclaves in SGX and trusted world 
in TrustZone).  The trusted hardware precludes the TTP
from altering the integrity of the contract code and examining
sensitive data.

\subsection{Contributions of this paper}
 Since the publication of the Bitcoin paper~\cite{Satoshi2008}, smart contracts have been discussed mainly within the context and confine of blockchains to which they seem to be intrinsically associated. As a result, the potential and practicality of smart contracts deployed on Trusted Third Parties (on--TTP) have  been largely underrated. The intention and contribution of this paper is to restore balance. In pursuit of this aim we discuss  the advantages and disadvantages of using on--blockchain and on--TTP smart contracts.
Our arguments are the results from the following exercise. As developers we have been asked to advise on the selection  of a smart contract technology to implement a car insurance application. To be able to  offer  a  well  supported and defined  recommendation we have conducted exploratory implementations (as opposed  to fully fledged) using on--chain and on--TTP smart contracts. In this paper we share our experience and give some insights into the technical problems that we have faced. We have learnt that some of them are easier  to  solve  with on--TTP smart contracts than with on--chain smart contracts. We acknowledge that a noticeable disadvantage of an on--TTP smart contract is its dependence on a single TTP---a design feature that is considered undesirable and avoidable via the use of blockchains. However, we argue that in applications where the inconveniences of a TTP can be tolerated, on--TTP smart contracts might be a better alternative than their on--chain equivalents.
 
 The remain of this paper is organised as follows.  To
open the discussion, Section~\ref{motivatingscenario} 
introduces the hypothetical use case of a  car insurance application that we
have been asked to implement as a smart contract
application. Section~\ref{centralisedanddecentralised}
explains that smart contracts can be implemented either 
following centralised and decentralised architectures.
Section~\ref{onblockchainimplementation} discusses the
experience we have gained from the implementation of the
car insurance application using an on--chain smart
contract implemented in Solidity language to be deployed
on Ethereum. In Section~\ref{onTTPimplementation} we 
include complementary arguments gained from the 
implementation of the example, using an on--TTP
smart contract.
In Section~\ref{discussion}, we express our
personal views of smart contracts and mention
open research questions.  
To place this article within a broader context, Section~\ref{relatedwork}
provides a summary of pioneering articles on 
smart contracts and recent publications that are related to
this article. Consequently, we draw conclusions in Section~\ref{conclusions}.

\section{Motivating scenario: car insurance application}
\label{motivatingscenario}
\begin{figure}[!t]
	\centering
	\includegraphics[width=0.65\columnwidth]{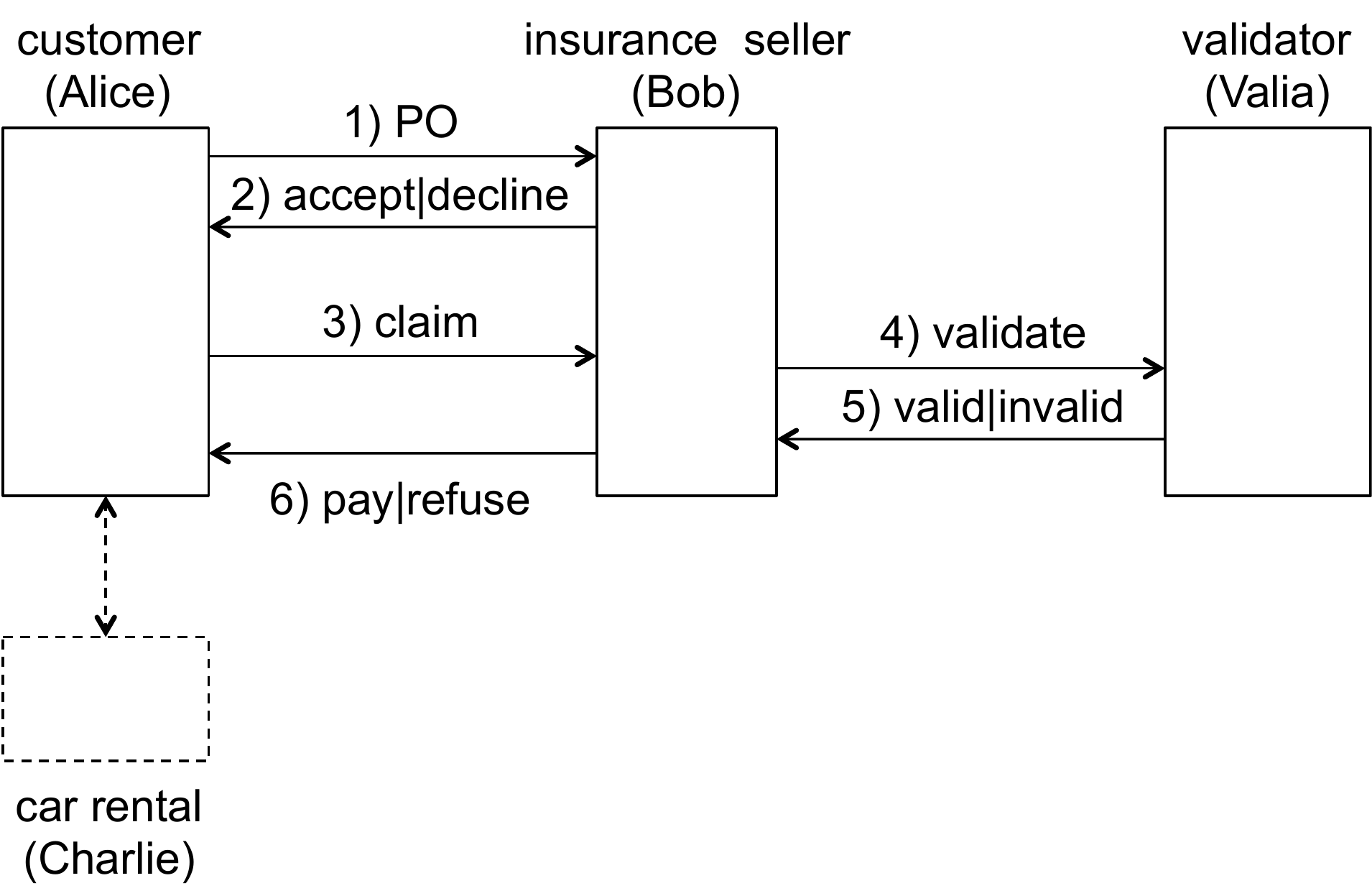}
	\caption{Car insurance application.}
	\label{fig:carinsuranceapplication}
\end{figure}

A high level view of the hypothetical use case car insurance scenario is shown in Fig.~\ref{fig:carinsuranceapplication}. It involves three participants, namely, a 
\emph{customer} called Alice, a car insurance company called \emph{insurance seller} run by Bob and a claim \emph{validator} (Valia). The scenario assumes that Alice has rented a car from a 
\emph{car rental} company run by Charlie and that she is now interested in purchasing an insurance policy. A more comprehensive
implementation would involve the car rental, however, to simplify the task, 
we have left that out of the current implementation. 

The figure shows the execution of the six main operations:
\begin{enumerate}
\item Alice submits a Purchase Order (PO) request to Bob 
      to purchase her insurance policy.
 \begin{enumerate}
   \item The PO is a document that includes the payment to cover 
    insurance premium, car registration details, Alice's personal 
    details and Alice's preferences regarding the  storage of her personal details.
   \item Bob stores Alice's personal details in
      accordance with Alice's preferences, she is entitled to choose
      from two alternatives:
      \begin{itemize}
          \item \textbf{delete:} Bob's insurance company is expected
            to delete Alice's personal details within  one
            day of declining to sell the insurance or 
            one day after the expiration of the insurance certificate.
         \item \textbf{store:} Alice grants permission to the insurance
           company to store her details indefinitely unless
           they are notified by Alice to delete them. In the
           latter case, the insurance company is expected
           to delete Alice's details within 14 days of
           receiving Alice's request.
      \end{itemize}
  \end{enumerate}
\item Bob uses his discretion (for example, on the basis of
      the registration number) to \emph{accept} or 
      \emph{decline} Alice's PO, within 10 min of submission
      of the PO.
    \begin{enumerate}
      \item if Bob accepts, he proceeds to collect Alice's payment
       and declare the insurance policy in force.
    \end{enumerate}
\item In the event of an accident, Alice submits an insurance
      \emph{claim} to Bob.

\item Bob consults with his validator (Valia) to ascertain the legality of 
      the claim (see \emph{validation} operation).
\item Valia declares the claim either \emph{valid} or \emph{invalid}.     
\item On the basis of Valia's verdict, Bob proceeds to either 
      \emph{pay} or \emph{refuse} the claim within 24 hours of the submission 
      of the insurance claim. 
\end{enumerate}

\begin{figure}[!t]
	\centering
	\includegraphics[width=0.65\columnwidth]{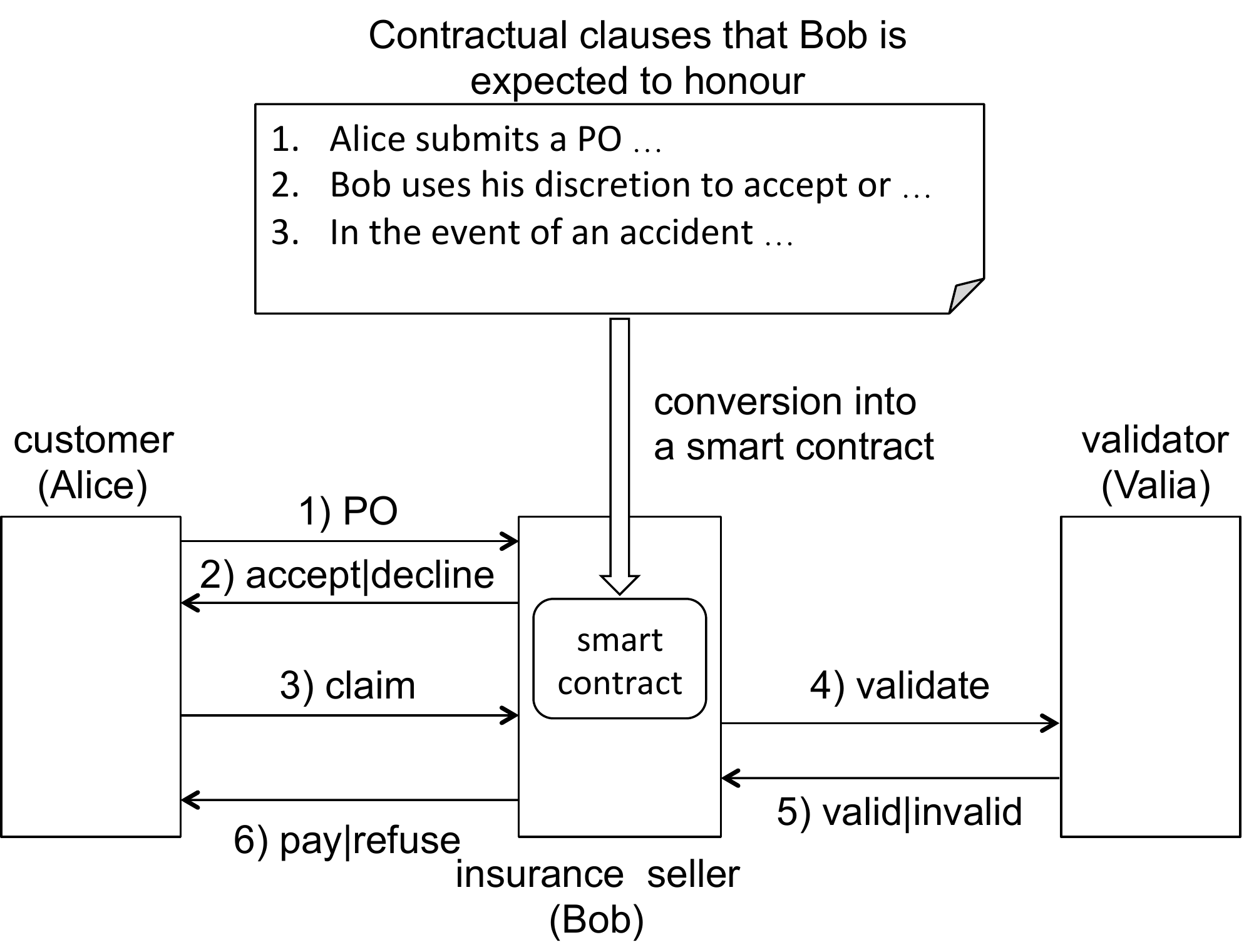}
	\caption{Architecture of the car insurance application.}
	\label{fig:carinsurancesmartcontractarch}
\end{figure}

An abstract view of a smart contract--centric architecture that
can be used for implementing the car insurance application is shown in Fig.~\ref{fig:carinsurancesmartcontractarch}.

 The \emph{smart contract} box within the insurance seller is
 the result of converting the legal clauses shown above into 
 executable code, for example, into Solidity or any other
 computer language including Java and Python.
 The smart contract is in control of the insurance process; as such,
 it is responsible for executing the contractual operations when 
 certain conditions are met and the execution of the operation
 is triggered. For example, if Alice stipulates that her 
 personal details shall be deleted within a day of failing
 to purchase an insurance policy, the smart contract is 
 responsible for deleting Alice's personal details within a day. 
 Likewise, in the event of a successful purchase followed by a 
 valid claim, the smart contract is responsible for triggering
 the execution of the pay operation to pay the relevant 
 determined compensation to Alice within 24 
 hours of the submission of the claim.


Let us assume that the smart contract support only the six operations 
mentioned in the contractual clauses listed above: PO, accept, decline, 
claim, etc. This level of detail is enough to present our arguments. For 
the sake of simplicity, the smart contract does not account for insurance 
cancellation and other operations that a more realistic application would 
include and which would be more complex.

Notice that the smart contract is responsible for managing  the entire 
insurance process which includes the relevant provisions  concerning Alice's 
personal details. This comes   into force when Bob collects Alice's personal 
details and completes when Bob deletes Alice's personal details from his 
records.

 \section{Centralised and decentralised smart contracts}
 \label{centralisedanddecentralised}
  The hypothetical example of smart contract of the motivating
  scenario is rather simple
 and can be modelled as a conventional Finite State Machine
 (FSM) and converted into executable code. Having decided 
 to use a FSM model, developer needs to make a crucial decision
 at this point: where to deploy the smart contract, as explained in
 Section~\ref{smartcontractdeployment}.  
 
 In general, depending on the number of instances (replicas) deployed 
 of the smart contract and on who is in control of the
 instances, we distinguish between two alternatives:
 centralised and decentralised (see Fig.~\ref{fig:centralisedanddecentralisedcontract}).
 \emph{A} and \emph{B} are business partners, for example, Alice and 
the insurance company of our contract example. \emph{SC} is the corresponding 
smart contract. \emph{op} stands for operation executed against 
\emph{SC} (for example, PO), \emph{rp} is the corresponding response 
(for example, PO approved). \emph{TTP node} is a node under the 
control of a TTP. 

\subsection{Centralised deployment}
In a centralised deployment, we instantiate only a single instance 
of the smart contract of a node controlled by a TTP (see
Fig.~\ref{fig:centralisedanddecentralisedcontract}--a)).
Since the TTP node is not part of the blockchain,
this deployment is referred to as an \textbf{off--chain
deployment} and as an \textbf{on--TTP deployment} to emphasise
the fact that the TTP plays the role of a central
authority and is in control of the application.

\subsection{Decentralised deployment}
\begin{figure}[]  
	\centering
	\includegraphics[width=0.85\columnwidth]{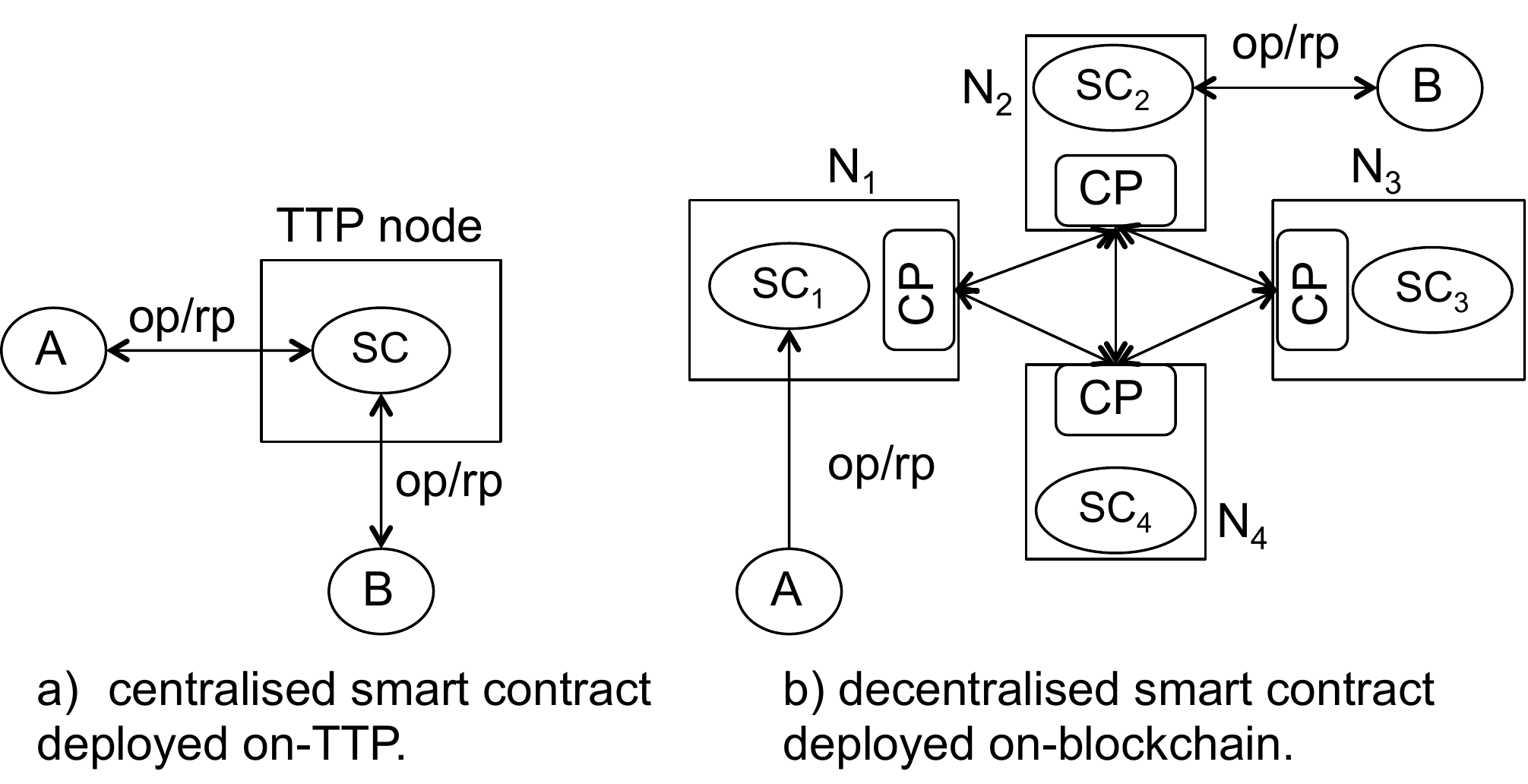}
	\caption{Centralised (on--TTP) and decentralised (on--chain) smart contract.}
	\label{fig:centralisedanddecentralisedcontract}
\end{figure}

In a decentralised deployment, we instantiate $N\geq2$ identical 
instances of the smart contract.
 Each of the $N$ instances keeps a local copy of the state of the contract, 
 consequently this deployment needs sophisticated middleware support 
 to synchronise the copies. The deployment shown in 
 Fig.~\ref{fig:centralisedanddecentralisedcontract}--b) relies on
the middleware support offered by a conventional blockchain. In the 
figure, four identical replicas of the smart contract
(${SC}_1,\dots,{SC}_4$) are deployed on a 
blockchain. $N_1,\dots,N_4$ are untrusted nodes, each operated
by an independent owner and members
of the blockchain network, for example, Ethereum.
This deployment is referred to as \textbf{on--chain}; it is
also referred to as \textbf{decentralised} to emphasise its
peer--to--peer model of interaction between the parties
and its distributed computing and the storage and maintenance
of complete copies of all transactions.

 In this approach, \emph{A} and \emph{B} are free to place their operation against 
 any of the instances.
The price that the decentralised approach pays for getting 
rid of the TTP is that the untrusted nodes must run a 
consensus protocol (represented by \emph{CP} ) such as
Proof--of--work, to verify that 
a given operation has been 
executed correctly, and to keep the states of 
${SC}_1,\dots,{SC}_4$ identical.

It is important to note that decentralisation is a spectrum that can range 
from partial decentralisation to full decentralisation of a particular network. 
Furthermore, the level of decentralisation must also naturally take into 
account a variety of factors including both soft and hard factors. Hard factors 
may include the actual architecture of the network as well as the relevant 
consensus mechanism algorithms for that blockchain, as well as the 
concentration of nodes by node operators and computing power of miners 
and mining pools. Soft factors on the other hand include the way and manner 
with which upgrades and changes to the software takes place, the relevant 
concentration of developers, thought leadership of notable developers or 
experts, and other potential factors that may influence the trajectory of the 
network and its development, be it positive or negative in pursuit of particular 
objectives. 

It is also crucial to distinguish between
distributed from decentralised. 
Distributed refers primarily to the hardware and software infrastructure. It determines where the computation takes place (on a single or several computers) and how the records of the transactions are stored (a single copy on a single ledger or on several copies on several ledgers). On the other hand, decentralisation refers to the ownership
of the nodes of the infrastructure: are the nodes under the control of
a central authority or run by independent owners?
A blockchain may be decentralised but not distributed and distributed but not decentralised. For example, Fig.~\ref{fig:centralisedanddecentralisedcontract}--b) would be a distributed but centralised blockchain if $N_1,\dots,N_4$ 
were under the control of a single authority. Depending on the degree
of control that a central authority exercises on the underlying 
infrastructure, there are different levels of centralisation and 
decentralisation. Fig.~\ref{fig:centralisedanddecentralisedcontract} illustrates
the two ends of the spectrum. The idea of fixed and rotating committee
decentralisation is an example of how decentralisation can be gradually relaxed~\cite{Gorbunov2018}.

\vspace{0.5cm}
 As developers we have considered the two alternatives to
 implement the car insurance application. In the following sections we discuss our experience.

 \section{on--chain implementation}
 \label{onblockchainimplementation}
 With this approach, the developer has the middleware services offered by 
 the blockchain (see Section~\ref{blockchain}) at his disposal. Unfortunately, along the benefits, the blockchain brings several 
 complexities and engineering problems that are extremely hard to solve. 
 To illustrate the point, let us take Ethereum as the implementation 
 blockchain platform with the assumption that other blockchains 
 that follow the conventional blockchain model will bring similar
 advantages and technical problems.

 \subsection{Implementation and code}
We have published the Solidity code at Git~\cite{carInsuranceExample2019}.
The core of the code is the \emph{contract CarInsurancePolicy\{...\}}
smart contract. Essentially it implements a finite state machine
that keeps track of the set of states $S=\{CREATED, APPROVED, CLAIM\_MADE, \dots \}$ of the policy purchased by the customer or policy holder. The finite state machine
progresses from ${PolicyState}_i$ to ${PolicyState}_{i+1}$
in response to actions executed by the customer (represented by
Alice), underwriter (represented by Bob the insurance seller) 
or validator (represented by Valia).

To enable the application to handle several insurance policies
instances simultaneously, we have implemented
the \emph{contract CarInsurancePolicyManager\{...\}}. This is
the smart contract and is responsible for creating and managing
policy instances. The current implementation can handle only 
a single underwriter 
and a single validator. A more
realistic implementation would need to handle a list of 
several underwriters and validators. 

Alice interacts with the smart contract as follows:

\begin{enumerate}
    \item Alice is assumed to have a GUI integrated with a software
     capable of creating Ethereum Wallets such as 
     Metamask~\cite{metamask}.

   \item Alice uses her GUI to place insurance purchase orders which
    include her personal details. The submission
    of a purchase order causes the execution of the 
    \emph{createPolicy(carReg, msg.sender, ...)} function of the 
    \emph{CarInsurancePolicyManager} smart contract. As a result,
     the smart contract creates and registers a new policy instance. 
     The function notifies Alice's GUI about the outcome by means of the
    \emph{NEW\_POLICY} event.
    
    \item Alice uses her GUI to execute actions against her 
     policy. For example, when the state machine is in state
     \emph{APPROVED}, Alice is able to make a claim 
     which would result in the execution of 
     \emph{function makeClaim( ... )} and progresses from $APPROVED$ 
     to $CLAIM\_MADE$ state.
\end{enumerate}

The underwriter and validator interact with the \emph{CarInsurancePolicyManager} smart contract similarly.

There are several tools that a developer can use for deploying and visualising Ethereum smart contracts. In this occasion we have opted for the following technologies:

 \begin{enumerate}
    \item We use the Truffle deployment framework which is currently 
     the de--facto standard. Truffle offers several
     services, including, compilation of Solidity into EVM code,
     linking, deployment of the contracts and binary 
     management~\cite{truffleHome}.
    \item We deploy the smart contract from a Parity Ethereum client---a 
    Rust~\cite{RustHomepage2018} implementation of Ethereum client protocol. 
    There are several implementations such as geth (a Go~\cite{GolangHomepage2018} implementation)
      that would have worked as well~\cite{GethEth}. The only requirement
      is that the client is synchronised (in possession of the
      latest blocks) with the Ethereum
      blockchain. Synchronisation might take hours, depending on the 
      current status of the client. Another client alternative is 
      infura~\cite{infuraHome} which is essentially geth or Parity
      in the cloud.
    \item We use one of our Ethereum accounts to deploy the \emph{CarInsurancePolicyManager} smart contract. It
       has enough funds to cover the deployment
       fees (about 4.9 Milliether).
    \item The account is used by Truffle to sign the 
       Ethereum deployment transaction before broadcasting it through
       the Parity client (or geth client if geth is used).
    \item As can be seen at Etherscan~\cite{carInsuranceExampleEtherscan2019}
    the \emph{CarInsurancePolicyManager} smart contract  has been
    deployed at the 0xb3C66fA11af5b4975D74C654665A0b7E505b2bDe address.
 \end{enumerate}
  
We will focus our attention now on the discussion of the technical
issues that we have faced in this implementation.

 \subsection{Encryption issues} 
 As explained in Section~\ref{motivatingscenario}, a purchase
 order request includes car's registration number and
 personal details of the purchaser, for
 example personal details of Alice. In the example, this information is included, respectively,
 in the \emph{carReg} and \emph{msg.sender} parameter passed to
 \emph{function CreatePolicy(...)}. The function needs to store 
 the actual documents or a hash of them which references the
 actual documents, wherever stored.
 
       The encryption/decryption of Alice's documents is 
        cumbersome.  For security reasons, it is sensible
        to keep documents encrypted.  The problem is that 
        they need to be read--shared by several parties
        under different access policies.
        For example, the \emph{carReg} needs to be accessed by both 
        Bob and Valia. 
        If the documents are small, one can encrypt and store the actual documents on--chain. 
        However, once a document is broadcast to the public
        on a decentralised blockchain, revocation of access rights 
        becomes technically impossible. There is no systematic model for
        controlling access to encrypted documents stored on
        blockchains. Fully  decentralised access control is 
        impossible. An alternative however, is key sharing  but the 
        immutability of the  document makes the task much more difficult. 
        One could make the  argument that revocation of access rights is 
        against the principle of transparency in blockchains, yet in practical 
        applications access control policies are needed to protect both the 
        integrity of the data, and the personal data, for example in this case 
        that of Alice.

\subsection{GDPR compliance issues} 
Compliance with GDPR requirements is problematic. The indelible append--only 
model that underpins the blockchain conflicts with EU General Data Protection 
Regulations  (GDPR) which came into force in the UK in May 
2018~\cite{GDPR2018,GDPRinfo2018}. These regulations grant individual the 
right to request deletion of 
their personal  details under Article 17 through the right of an individuals to be 
forgotten. It follows that Alice's request to delete her personal details after failing  
to  buy  her  insurance  or  at  some  point  in the future is to be taken by the insurance 
company as a   legal requirement. Notice that  under  this  regulation  hashes  of 
personal would most likely be regarded as personal data. There is no practical 
solution to this problem in smart contracts deployed on--chains at present. 
Furthermore, depending on the nature of the activity, transaction or contract, compliance 
with other data privacy regime may be required depending on the location of the 
parties to the system, but also potentially engaging the restrictions related to 
international personal data transfer under GDPR.
Other GDPR concerns include the possibility for a public key, which is used to sign 
multiple different smart contracts, may lead to the identification of a particular individual. 
In such an instance, it is arguable that the public key would then be considered 
personal data as it leads to the identification of the owner. Moreover, assessing and identifying 
the relevant ''controller'' and ''processor'' under GDPR regulations may serve to be 
counterintuitive to the overall schema of using a decentralised blockchain and 
on--chain smart contract. The reasoning behind this is as follows, if personal 
data is entered into a block by an individual or business, they may likely 
be categorised as 
a ''controller'' for the purposes of GDPR unless it is only supporting 
the blockchain network, in which
case it would merely be a "processor" and which raises the concern that nodes and 
miners could each be considered "processors" if a public key, used to sign multiple 
smart contracts is considered personal data due to the fact that it references information 
which could lead to the identification of an individual.
  It is important to highlight that even if code were law, GDPR regulations would remain 
of crucial concern. Code currently is not considered law, and therefore any given smart 
contract, or code, is by definition referencing the relevant documentation and contracts 
which stipulate and define the relevant rights, duties, and obligations of the parties and 
are looking to mirror the relevant provisions. 
Naturally, this is relatively cyclical in the sense that in order for code to be considered law, 
the law must recognise it as such in the form of legislation or common law applied by 
the judges and the courts where smart contracts, or code, are an accepted contractual 
agreement, and which can be submitted as evidence, and from which rights, duties, 
and obligations arise. 
Nevertheless, there are instance in which code may likely satisfy the relevant tests in 
different jurisdictions for being fully enforceable contracts, given that in certain jurisdictions, 
a physical paper contract is not required for a contract to be formed, and therefore there 
are certainly instances where a smart contract alone would be a contract. In which case, 
documentation would not necessarily be referenced, but rather, the code would be the 
contract itself.
Yet, even if physical or digital documentation were not required, there would still be 
required code stipulating the relevant agreement between the parties and assigning 
rights and duties to the parties, which would, at least in the present context, require 
the identification of the parties to some degree, and the inclusion of certain personal 
data which would lead to the same data storage concerns and GDPR issues related 
to an individuals right to be forgotten and data qualifying as personal data due to it
 potentially leading to the identification of an individual.

   \subsection{Off--blockchain storage issues}
       Storing and retrieving documents from off-chain
       storage is extremely complex.
       To save on storage, it is advisable to store large documents
       off--chain and use the blockchain to store only pointers
       (for example, a hash) to the actual documents.
       The Inter Planetary File System(IPFS)~\cite{JuanBenet2014,ipfs}
       has been suggested by academic researchers as off--chain
       storage. In practice this solution is hard to implement. The 
       difficulties emerge from the complex P2P protocol 
       that underpins IPFS. The whole protocol needs to be deployed on
       all parties. IPFS not only bloats the size of
       the application, but also complicates the application's 
       internal message flow. Also, the P2P protocol of IPFS
       exacerbates the problem of shared access to documents
       and GDPR compliance. IPFS stores content--addressed immutable 
       objects in a decentralised manner, consequently, it is cumbersome
       to manipulate file access privileges or to delete documents
       to honour GDPR requests.
            
\subsection{Number of variables issues} 
 In Solidity, each function is allowed to have at most 16 local 
 variables. Such constraints are usually worked around by handcrafted data packing, or by splitting one function into multiple functions. 
 Annoyingly, very often the developer finds himself spending more time on 
 overcoming such constraints than on the implementation of the actual logic.

   \subsection{Gas cost issues} 
   In theory, and as stipulated in the Ethereum yellow 
   paper~\cite{EthereumYellowPaper2019}, a message transaction can include
   an array of an unlimited number of bytes to be used as
   input data in the message call. In practice, the length of this
   array is constrained by the gas cost and the block gas 
   limit. Recall that to prevent denial of service attacks,
   Ethereum imposes a gas fee on the execution of each 
   transaction which depends on the number of computation
   steps needed to complete the transaction and its
   length in bytes~\cite{EthereumYellowPaper2019}. Also,
   Ethereum imposes a limit on the 
   amount of gas that a block can consume, consequently, the gas allocated to
   a transaction cannot exceed the block gas limit. The block
   gas limit is determined dynamically. On the 28 of Feb 2019, 
   the block gas limit is 8,001 071 gas~\cite{EthreumGasLimitChart2018}.
   In accordance with~\cite{Cryptostats2018}, the average number of
   transactions per hour is 23150 while the average number of
    blocks per hours is 177. It follows that currently the average number
    of transactions per block is about 23150/177=130. This means that
    on average each transactions is allocated about 8000000/130=61538 gas. 
    To have transactions processed, it is mandatory not to 
    exceed the block gas limit. Also, to have transactions
    processed by miners without delay, it is advisable to 
    work with transactions of average gas requirements. A technical
    difficulty here is that, though there is an algorithm that
    miners can use for determining the block gas limit, these figures 
    are dynamic. For example, on the 27 of Jun 2017, the block 
    gas limit was 4711731.

   The problem we had in the implementation of the
   car insurance application is that the issue about
   the block gas limit can impact functions with moderate
   loops. For example, each iteration for--loop of the 
   \emph{function updateValidator} of the \emph{CarInsurancePolicyManager} contract costs 8156 gas. Given the 8M block gas limit, this function's maximum number of iterations is about 980.
   This also becomes the maximum number of contracts that the policy manager should create.
   Technically, such limit can be overcome by deploying multiple policy managers. 
   Such workaround introduces extra complexity. This discussion shows that the 
   problem of scalability of Ethereum is not only due to a) its consensus 
   algorithm and b) its on--chain storage but also c) its computational model.  
   The latter point c) has not received enough attention; yet in our experience, it is 
   harder to solve as it is inherent in the Ethereum Virtual Machine (VM).  
   Side--chains have been suggested for addressing  the  gas  cost problem. The technique has different variants but  the  main idea is to use an additional  channel  for  conducting  some transactions off the main blockchain to off load it from frequent transactions occurring on the main blockchain. It is also referred to as plasma and state--channel~\cite{Andrew2017}. However, this solution does not work when one is looking for an integrated solution. Side-chains only delegate the issue and complexity to the developer responsible for implementing the side--chain which includes the management of the interaction between the main blockchain and the side--chain. Another matter is that a side--chain does not have 
  the mechanisms for complying with GDPR regulations unless its 
  implementation departs from the indelible 
  append--only model which defeats the purpose of using a blockchain based smart contract.

\subsection{Off--blockchain interaction issues}
A well documented technical problem that afflicts blockchains is the difficulty 
to make on--chain smart contracts interoperate with off-chain  
components  such  as data and applications. For example, it is hard for 
an on--chain smart contract to send 
notification events to off--chain components to drive their execution~\cite{Gideon2016,Fan2016,Davidhamilton2018} or to read 
information from applications operating off--chain.

Notice that this issue manifests itself in Fig.~\ref{fig:carinsuranceapplication} 
where the smart contract needs to notify Alice's application about 
the outcomes of her claim (pay or refuse). A way of  getting  around 
this difficulty is  to  implement  a  polling  mechanism in Alice's 
application to learn about the outcome. A polling--based solution 
works well in simple applications like this insurance example 
without strict time constraints  to  identify   the outcome.
 Alice's application is free to poll at its own time, without the pressure of time constraints. However, in situations with strict time constrains, notification mechanisms would be much  more efficient.
Another manifestation of  the  problem  is  the  notification  of Alice's accidents to the smart contract. Ideally, these notifications should be done directly to  the  smart  contract,  for example, by  an  IoT  device  embedded  in  the  car.  This  is a hard problem  that  we  have  not  yet solved.  Some  of  the challenges have been discussed above, for example, how to make sure this notification will be 
mined in a reliable and timely way? 
In our current implementation we do not account for direct notification; 
we opted for a simple solution where Alice submits her claim to the 
smart contract, which in practice would then be required to be verified 
or endorsed as a legitimate claim.
Technologies that bridge the on--chain and off--chain infrastructures, referred to as oracles, have been suggested to address the problem (see for example~\cite{oraclizeHome}). Conceptually the idea of oracles is simple, however, in practice, oracles cause difficulties. 
One of the most attractive features   of using  on--chain  smart  
contracts  is  the  avoidance  of a centralised TTP to prevent the 
existence of a single point   of failure. Sadly, oracles sharply 
diminish this feature as they operate as TTPs themselves adding 
centralisation to a decentralised blockchain. Decentralised 
oracles have  been suggested for ameliorating the situation 
but at the cost of additional complexity and at present, inferior 
technical development.

\subsection{Data inconsistency issues}
Like all distributed applications, on--chain smart contracts are at risk of consuming inconsistent data unless they   are provided with remedy mechanisms like exception handling. Such mechanisms need to handle situations caused by  the consistency model followed by  the  blockchain  used  in  the implementation. If the blockchain follows the eventual consistency 
model~\cite{Decker2016} like Ethereum and Bitcoin, then the mechanisms need to take into consideration that transactions are not always mined, a transaction that has been tentatively committed might suddenly disappear when the blockchain is reorganised, upgraded, or forked and so on.
These mechanisms exist and can be encoded into the smart contract but they increase  the complexity and more importantly, they prevent the smart contract from consuming timely data which increases the risk of inaccurate near real--time data to execute upon, and which therefore may no longer be 
contractually irrelevant or even worse, it might grant, suspend or cancel the wrong  rights, duties or
obligations.  This is a well known problem in distributed systems that manifests in on--chain smart contracts as well. To be safe, the smart contract needs to wait until the risk of being impacted   by a blockchain reorganisation minimises. The problem is that in Bitcoin it can take as long as 10 minutes to see a transaction included in a block. This is only a tentative commitment that cannot be taken as a definite transaction confirmation because there is still a risk of a blockchain reorganisation that might take place a few minutes later and destroy the transaction, rendering it to have never occurred and would require it to be re--initiated.  Similarly, to be safe from chain reorganisation in Ethereum, one needs to wait until 20 to 30 blocks are included in the ledger, which corresponds approximately  to  4  to  5  minutes of waiting time. This transaction delay impacts the response time of the insurance application. Incidentally, to take Alice's payment as valid and issue her an insurance certificate, Bob needs to see at least Alice's payment  transaction  is  mined  and recorded in the blockchain. Yet, depending on network congestion, Alice's transaction may take hours to be included  in a block. Afterwards, there is always a risk that  the new block becomes orphaned which would force the claim process to start from the beginning. Chain reorganisation in Ethereum is explained  
 in~\cite{GavinWood2015}. With current practice, the estimation is that the application needs to 
 wait for about 30 min to use the transaction safely which is too long for current VISA 
 standards.

\subsection{Nonce issues} 
The nonce is one of the attributes of an external account and is a monotonically increasing scalar equal to the number of transactions sent from the account~\cite{EthereumYellowPaper2019}. The nonce is used by
Ethereum in order to  prevent the execution of transactions more than once (transaction replay). Least understood is the fact that the nonce imposes a FOFI (First-Out-First-In) processing discipline on transactions out from a given account: transactions sent from a given account are included in blocks in the order of sending. The problem is that in some situations, this FOFI discipline brings unnecessary constrains. Imagine that in the car insurance example, Valia decides to send a transaction to approve Alice's claim, followed by a transaction to approve Clare's claim. In this example, Clare's transaction will be considered for inclusion in a block only after the inclusion of Alice's transaction. FOFI brings simplicity in programming and is useful in some situations. However, FOFI inhibits the use of gas price as a prioritisation mechanism within a single account. When the FOFI discipline is more of a burden than a contribution, the developer can resort to engineering solutions. For example, the developers can provide Valia with multiple private keys to break the FOFI order of her transactions. However, this not only over--complicates the implementation, but also increases the potential attack vector. The situation is further worsened by the data consistency issues discussed above.

\section{On--TTP implementation}
\label{onTTPimplementation}
The car insurance application can be implemented on the
basis of a smart contract deployed on--TTP. 
A TTP brings simplicity and governance. 
\begin{itemize}
 \item{\textbf{simplicity:}} A TTP obviates the need of heavy weight 
distributed algorithms which results in
low overhead. For example, a TTP--based smart contract
does not need to execute consensus over the next state of  
the smart contract or involve miners or verifiers. Consequently,
on--TTP smart contracts do not suffer from the lack of scalability
and performance issues that afflict blockchains seen by
the wait time previously discussed~\cite{Trent2016}.

\item{\textbf{governance:}} A TTP can act as a central authority
 that enforces governance. It can play a crucial role in dispute resolution situations 
 which may take the form of alternative dispute resolution mechanisms 
 such as mediation and arbitration, depending on the clauses stipulated 
 in the relevant contract but also in litigation.
 In each scenario the potential enforcement of contractual provisions, 
 the granting of an award, injunctions, damages and other forms may 
 necessitate the need for a central authority to comply with the necessary 
 requirements. Disputes in contractual agreements are not as rare as 
 one would expect. We resume the discussion of this point  in
 Section~\ref{discussion},  for  the time being  imagine  that  an  
 abrupt  termination  of  an Ethereum smart contract triggers a  
 dispute  between the contracting parties. The followings questions arise, 
 who will have enough authority and evidence to arbitrate the 
 dispute? How will the dispute be arbitrated or litigated? And how 
 and in what way does enforcement of an award or a judgement 
 take place? The existence of a TTP that has a global view
 of the interaction and collects non--repudiable records makes
 dispute resolution simpler. In contrast, resolution of disputes
 that emerge from on--chain smart contracts where no TTP
 exists are far more complicated.
  \end{itemize}
 
   Let us examine how the inclusion of a TTP prevents 
  or simplifies the technical difficulties that plague 
 on--chain smart contracts as discussed in 
 Section~\ref{onblockchainimplementation}.

\subsection{Implementation}
There are different technologies that can be used for conducting
the implementation. To conduct our analysis we use 
a Contract Compliance Checker (CCC) tool that we have
developed in previous projects for the enforcement of smart 
contracts~\cite{MolinaTSC2011}. 
We have used the CCC in the implementation of several
smart contract applications (see for example~\cite{Solaiman2015})
following the architecture shown in Fig.~\ref{onTTPimplementation}.
The CCC is an open source tool and available from the
conch git repository along with its user's guide~\cite{conch}.
It is implemented in Java and as shown in the figure, consists of
three layers (presentation, logic and data layer) that enable
developers to deploy it as a conventional web server.

\begin{figure}[!t]
	\centering
	\includegraphics[width=0.98\columnwidth]{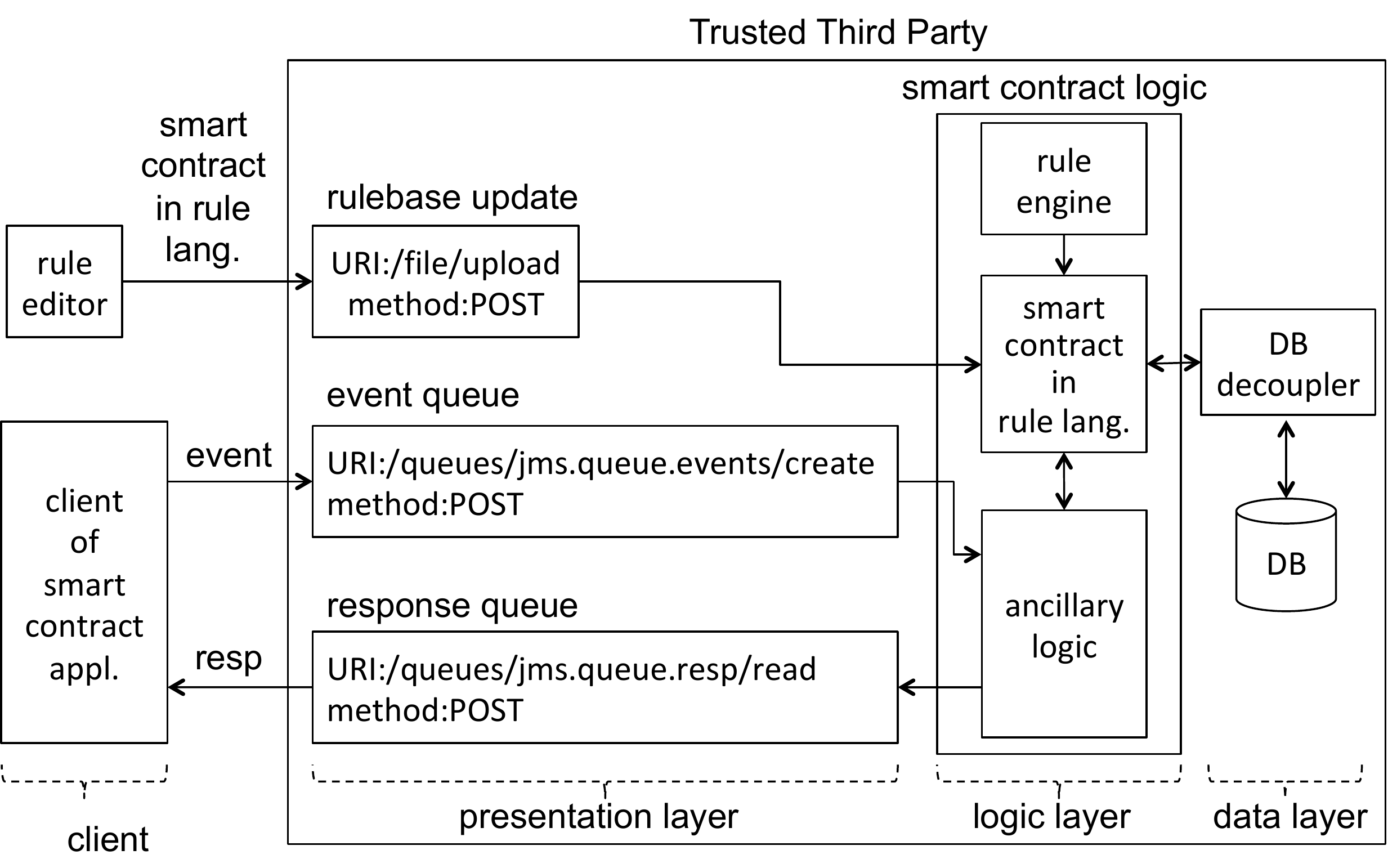}
	\caption{on--TTP implementation architecture.}
	\label{fig:onTTPimplementation}
\end{figure}

The smart contract logic is the core of the CCC and
includes the smart contract coded in the Drools
language supported by Jboss~\cite{Drools5.3.0UserGuide},
a \emph{rule engine} and \emph{ancillary 
logic}. They implement a FSM that grants and removes rights, 
obligations (duties) and prohibitions to the contracting parties as the 
execution of the contract progresses. The data base (\emph{DB})
is used for storing the history of the contractual operations 
and can be deployed locally or on a cloud provider.
To enforce a business contract with the CCC, the developer 
i) uses a conventional editor (\emph{rule editor}) to
write the contract and ii) deploys (with the assistance of the
\emph{rulebase update}) the
file (\emph{smart contract in rule lang.}) to 
the \emph{smart contract logic} where the \emph{rule engine} 
takes it as its rule base.

In this exercise, the commercial contract is the contract
discussed in the motivation scenario of 
Section~\ref{motivatingscenario}; therefore, the clients
to the smart contracts are Alice, Bob and Valia who
execute operations against the smart contract. The
clients (only one is shown in the figure) represent 
operations as events (\emph{event}) 
and queue them in the \emph{event queue}. The events are
collected by the \emph{ancillary logic} and used by the
\emph{rule engine} to trigger rules that determine
if a given operation is contract compliant or non--contract
compliant. The response (\emph{resp}) can be sent to any party 
interested in the verdict, including the client. Alternatively, the 
response can be collected from a queue by the parties. Normally the execution of an arbitrary
operation is associated to the verdict.
 A typical commercial contract results in many declarative 
 event--condition--action rules which may be contingent on one 
 another as well as otherwise directly or indirectly tied. To conduct 
 our analysis, we have implemented only some of them, including the 
 rules that handle Alice's submission of her purchase and her payment. 
To conduct our analysis, we have implemented only some of
them, including the rules that handle Alice's submission 
of her purchase and her payment. They are available from the
\emph{examples\_contracts/carInsuranceContract.drl} file
of the conch repository~\cite{conch}.

We will show the rule that deals with the placement
of the PO and use it to explain our arguments.

{\scriptsize
\begin{verbatim}
00# Rule 1: 
01 #  handles the buyer's right to submit a PO request.
02 rule "PO Request Received"
03  when
04   # Verify type of event, originator, and responder
05   $e: Event(type=="POREQ", originator=="buyer", 
06       responder=="seller", status=="success")
07       eval(ropBuyer.matchesRights(POrequest))
08  then
09   # Remove buyer's right to place another PO Requests
10   ropBuyer.removeRight(POrequest, seller);
11   # Add obligation to accept or reject PO to seller
12   BusinessOperation[] bos = {POconfirm, POreject};
13   ropSeller.addObligation("ReactToPO Req",bos,buyer,60,2);
14   System.out.println("*PO Req Received rule triggered");
15   #
16   # logic to process a contract compliant PO
17   # 
18   responder.setContractCompliant(true);
19 end
\end{verbatim}
}
Lines starting with \# are comments.
Line 02 specifies the name of the rule, \emph{PO Request Received}.
Line 05 indicates that the rule will be triggered by an \emph{POREQ}
event, that is, when a purchase order request is executed. Line 
07 determines if the event \emph{POREQ} is
contract compliant or not. Lines 08--18 are executed only
when the event is found to be contract compliant. An \emph{else}
block can be added to the rule to cover non--contract compliance 
outcomes or alternatively, a separate rule can be included
to handle that---it is only a matter the programmer's preferences. 
Line 16 is meant to be replaced by the actual logic to process
the PO.

Once the smart contract is deployed, it is able to receive  
events (for example \emph{POREQ}) from clients submitted 
as RESTful messages. For example, Alice's application
can produce the following code and send it as a RESTful message
to the smart contract to notify the submission of a 
purchase order:

\label{restfulmsg}

{\scriptsize
\begin{verbatim}
  <event>
    <originator>seller</originator>
    <responder>buyer</responder>
    <type>POREQ</type>
    <status>success</status>
  </event>
\end{verbatim}
}
 
 In addition to the type of the event (\emph{POREQ}), the
 code indicates who the originator and responder
are. The status of the event indicates that the execution 
of the operation was successful as opposed to failure, 
this issue fall outside the scope of this paper. It is discussed in~\cite{MolinaTSC2011}.

Similarly, the smart contract can format its verdicts 
about contract compliance as shown in the following code 
and convert them into RESTful  messages.

{\scriptsize
 \begin{verbatim}
  <result>
   <contractcompliance>true</contractcompliance>
  </result>
\end{verbatim}
}

\subsection{Encryption issues}
The centralised storage that the database provides simplifies 
the issue about storing Alice's documents including her personal 
details. For instance, in the rule shown above, the developer can 
include access control mechanisms in line 16. To be pragmatic, 
the developer can resort to conventional and well understood 
server mediated access control mechanisms 
such as RBAC~\cite{Myong2001}. Alternatively, to add more security 
at the expense of simplicity, he can opt for attribute--based encryption 
mechanisms that enable to the  sharing of encrypted documents 
between several users under different policies and various 
parties~\cite{Bethencourt2007}.

\subsection{GDPR compliance issues}
 The inclusion of a conventional data base simplifies the
 issue about compliance with GDPR May 2017 regulations. 
 Upon Alice's request, the on--TTP smart contract can locate 
 Alice's documents in the data base and delete them.

 \subsection{Transaction block size issues}
  This is an issue of standardisation of smart
  contract hosted on TTPs; to the best of our knowledge,
  there are no standards. We believe that with the maturity
  of smart contract technology, smart contract hosting will
  be offered on the cloud as a service.
   
 \subsection{Storage issues}  
 The database offers conventional storage which is uniform, with 
 no distinction between expensive on--chain storage and cheap  
 off--chain  storage.  Thus the  developer  is  free   to program 
 without worrying about the cost of the storage consumed by
  each transaction.
  
\subsection{Gas cost issues}
The issue about the gas cost hindering complex business
logic loses relevance. Yet, the TTP party might charge for 
hosting the smart contract, or by the various outcomes, 
transactions or other business models. We expect the charge 
to be comparatively small as an on--TTP smart contract is 
simpler and does not involve consensus, mining and multiple 
verifications  of transactions.

\subsection{Off--blockchain interaction issues}
The single instance  of  the  smart  contract  can  call  
APIs  of applications directly to retrieve data without 
any risk of processing inconsistent and untimely data like 
in the on--chain approach~\cite{Gideon2016,Fan2016}. 
This point is illustrated by the code of the 
\emph{POREQ} event and the response shown in
 page~\pageref{restfulmsg}. The exchange of 
information between the smart contract and the
clients is conducted by means of conventional
RESTful messages. More details about the  operation
of the CCC as a web server are discussed 
in~\cite{Solaiman2015}.

\subsection{Data inconsistency issues}
The actual on--TTP smart contract can be regarded as a centralised
entity that has a global view  of the interaction that takes place
between the contracting parties over conventional Internet
channels. Its central logical position preserves it from the data 
inconsistency issues that threaten distributed applications.

\section{Discussion}
\label{discussion}
The central argument of this paper is that smart
contracts are a technology in their own right and not
intrinsically related to blockchains or distributed
ledger technologies. To appreciate
the point, it is perhaps useful to bear in mind that
blockchains are an aggregation of several independent 
technologies that Satoshi creatively integrated together to 
build the Bitcoin cryptocurrency platform in 2008~\cite{Satoshi2008}.
These technologies predate Bitcoin and naturally
have their own independent existence. Smart contracts are only one 
of them. Smart contracts were aggregated 
in the creation of Bitcoin to control (prevent the execution of illegal transaction) 
the exchange of BTCs. To reinforce this point, let us mention that 
peer--to--peer models  like the one that
underpins the decentralised model of Bitcoin and
similar blockchains have been known for decades~\cite{Viveki2003}. Likewise, indelible append only file systems like the database that blockchain uses for storing records about transactions
have also been known for decades, see for example IPFS~\cite{ipfs,JuanBenet2014}. The proof of work protocol that Bitcoin uses to prevent miners from conducting sybil attacks is another component technology; it has been known since at least 2002~\cite{Douceur2002}. The point  we  are  making  here  is  that the technologies used to compose a blockchain (for example, Bitcoin blockchain)
 can be used independently when the inclusion of the whole 
blockchain is more of a burden that a benefit.
 
 Disputes may emerge generally around the contractual arrangement 
 between the parties but it may also emerge in smart contract
 applications from errors introduced at design, implementation and
 execution time. Omission and implied terms are example of
 design time errors that might lead to disputes. Omission errors are 
 hard to avoid because it is hard or unpractical to anticipate and account 
 for absolutely for all possible factors that might impact a contract. 
 Disputes might emerge from errors introduced at implementation time 
 by programmers when they translate from contract text into 
 code ---there is a risk of misinterpretation of the contract text. 
 At execution time,  errors in the functionality of  the  infrastructure  
 used for executing  the smart contract might lead to disputes. In
 An insightful discussion about dispute
 resolution is provided by Markus Kaulart's presentation~\cite{codelegit2019}. 
 
To appreciate how the inclusion of a TTP simplifies dispute
resolutions it might help to have a look at the protocols
suggested for dealing with disputes in state channels
used for conducting off--blockchain payments~\cite{Stefan2019,Andrew2017}. Since they rely on smart contracts deployed on blockchains to deal
with disputes, they are complex and might incur expenses. 
 Another observation is that these protocols are meant to solve disputes mechanically; this is sensible but in our view, works only for simple disputes; complex disputes require a certain degree of human judgement, especially when issues of interpretation, rectification and implied terms are being considered, as well as in the assessment and granting of a particular quantum of damages or the relevant remedy to be granted, for example in the form of an award granted by an arbitral tribunal if a dispute moves to arbitration. 
 Furthermore, other concerns that would need to be addressed in more complex commercial scenarios include the applicability of the laws of multiple jurisdictions, in particular in cross-border commercial scenarios where the governing applicable law of the smart contract must be identified on a decentralised system, or where it has been stipulated in the contract and country X's laws are the governing laws over the smart contract yet the activity occurs in another jurisdiction, country Y's laws may also apply to the transaction or activity.  
 As an example, let us take an scenario where a supermarket 
offers two hours of free parking to customers and fines customers 
that exceed the time limit; in legal parlance, 
the customer pays a penalty for the breach of a contract. A smart  
contract  can  be  used for detecting arrival and departure times and fining customers automatically. However, the developer needs to account for borderline situations that are likely to be disputed and cancelled. For example, customers that overstayed for just  a few seconds are likely to complain; so will customers that cannot cross the exit barrier in time because of queues caused by emergency vehicles or other potential factors that the parking lot owner, in this scenario the supermarket, is responsible for.

In this situations it helps to have a centralised and 
authoritative TTP to arbitrate, perhaps with human 
assistance. Non-surprisingly, TTP arbitration is widely used in online businesses, and the last several years have seen a surge in online alternative dispute resolution mechanisms. See for example, the dispute
resolution mechanisms used by PayPal~\cite{PayPalDisputeResolution},
Amazon~\cite{AmazonPayDisputeResolution} and
Alipay~\cite{Rongbing2015}.
Having said that, note that some commercial institutions
aimed at dispute resolutions in blockchains are 
emerging ~\cite{Clement2018,codelegit2019}. An overview of the legal and 
technical challenges that they need to face are discussed 
in~\cite{Mateja2018}.   

The actual benefits of using blockchains against existing
and well established technologies like conventional
databases and, more importantly, about the practicality of fully
decentralised models (no TTPs at all) have been questioned 
by several authors~\cite{Kieron2017,Jimmy2018}. We are of the
opinion that on--chain smart contracts can help but they
cannot solve the whole problem without the assistance of
TTPs. The question that arises here is when is the assistance 
of  an  on--chain  smart contract strictly necessary and when is an 
on--TTP smart contract enough.   To answer the question the 
designer needs to identify the  unique feature or features (for example, 
trust, transparency, reliability, etc.) that only a blockchain can 
contribute and whether those required features necessitate 
the use of a blockchain that is decentralised. A frequently  mentioned  
attribute  of  blockchains  like  Bitcoin  is that they welcome people 
rejected by the conventional banking system to conduct transactions 
and engage directly in peer-to-peer activities. This is true, but emerging payment systems like Alipay~\cite{Rongbing2015} are another  alternative.  Thus in  the car insurance example of 
Section~\ref{motivatingscenario} implemented  following 
the on--TTP approach, Alice can use Alipay to pay. As a counterexample, we agree with the use of blockchain to  provide trust and transparency in applications run in countries afflicted by corruption for example. That is, in places where the use  of TTPs is unrealistic to provide accurate and trusted management of the hosting of a smart contract~\cite{Cubas2018}.
It is worth clarifying  that  in  addition  to  on--chain  and on--TTP 
there are other alternatives for deploying smart 
contracts~\cite{MolinaSOCA2011}. Among them is the 
on--Provider 
deployment where the smart contract is deployed within the 
premises of   the provider. This model is the norm for the 
traditional utility providers such as water, gas, electricity, 
broadband and cloud providers. With this deployment, the 
customer has no choice but to take whatever the verdict 
(for example about resource consumption and billing) of the 
smart contract as trustworthy. 
However, trusted hardware can be used to ameliorate the problem;
for example, the consumer can take advantage of the remote 
attestation facilities provided by trusted hardware to verify 
the authenticity and integrity of the contract code 
that the provider deploys to take the measurements 
of the consumption. See Section~\ref{onTTPdeployment}.

\section{Related work}
\label{relatedwork}
A pioneer of smart contracts is Naftaly Minsky from Rutgers 
University who started studying this topic around 1985 under 
the name of Law Governed Interaction (LGI)~\cite{Minsky1991}. 
The need to implement mechanism for formally enforcing contractual obligations and permissions
like bank loan applications is discussed in 1985 in an early 
publication~\cite{Minsky1985}. The implementation of \emph{moses},
a system for enforcing a shared set of rules (the Law) between several distributed agents is discussed in~\cite{MinskySerban2005}. Another pioneering paper in smart contracts is due to 
Lindsay Marshall from University of Newcastle. It suggests the use of contract objects for
modelling management policies~\cite{Lindsay1993}.

A model for enforcing state obligations involved in service delivery 
with strict service level agreements (SLAs) suggested by IBM
is discussed in ~\cite{ludwig2003soa}. An implementation of a
system aimed at monitoring and enforcement of state 
obligations is the Heimdhal engine~\cite{PedroGama2006}. Heimdal is
centralised and deployed within the premises of the service provider.
In~\cite{MolinaTSC2011}, the authors discuss the Contract Compliance 
Checker (CCC); a TTP--based tool for enforcing contractual right, obligations and prohibitions expressed as declarative Event Conditions Action rules. An approach somewhat similar to the contract compliance checker discussed in~\cite{MolinaTSC2011} is presented in~\cite{perringodart}, where permissions, obligations and prohibitions are mapped into ECA (even--condition--actions). An 
executable contract becomes a set of ECA rules deployed within a 
trusted third party and placed between the two business partners 
to drive the interaction between the two business partners.  

Some of the technical difficulties   that we have
discussed in Section~\ref{onblockchainimplementation} have
been identified and documented by developers. 
The issue about the lack of scalability of blockchains
has received substantial attention. For example,
consensus algorithms that are more efficient than
the probabilistic proof of work (PoW)~\cite{KenjiSaito2016} 
used originally by Bitcoin, have
been suggested~\cite{Marko2015,Shebar2017}. Other authors have 
suggested the alteration of the blockchain architecture.
For example, one can augment the main blockchain
with a secondary or tertiary infrastructure that is used for 
executing some of the transactions rather than executing
them on the main blockchain. This idea was originally suggested 
in the Lighting Network~\cite{Joseph2016} and expanded
by other authors in different modalities and under
different names such as state-channels, 
side--channels~\cite{Andrew2017} plasma and
hybrid architectures.

The key idea in hybrid architectures is that on--chain
 and off--blockchain smart contracts do not preclude each 
 other; on the contrary, on--chain and
 off--chain smart contracts can be combined together
 to take advantage of both approaches. 
The advantages and disadvantages of this approach
are discussed in~\cite{CarlosIoannisTurin2018,Eberhardt2017}.
There are different variants of the idea. In~\cite{CarlosIoannisSC22018} the 
smart contract is split so that 
 some of its clauses are enforced by an on--chain components 
 while the rest of the clauses are enforced by a smart contract 
 deployed on a TTP depending on the type of transaction and the 
 type of clauses. Two additional examples of actual 
 implementations are discussed in~\cite{GuyZyskind_enigma2015,Florian2016}.

Another
technique suggested for addressing the scalability
problems of blockchains, that involves alteration
of the blockchain architecture is sharding. The
central idea of sharding is parallelisation with
subsequent cross--shard consensus: several
transactions are executed in parallel by different
independent groups of nodes (called shards) selected 
from the group of conventional miners (see for example 
Zilliqa~\cite{Zilliqa2017}). To avoid the risk of
consuming inconsistent data one can replace consensus
algorithms that offer only eventual consistency
like the PoW algorithms with algorithms that offer stronger
consistency~\cite{Shebar2017}. Examples of such algorithms
are the classical Byzantine Fault--Tolerant 
(BFT)~\cite{Marko2015} that are capable of delivering
guaranteed consistency.
A common feature of the above techniques is that they
are aimed at circumventing the drawbacks of 
blockchains. In our view, these technique should
be used when the benefits that the blockchain brings
outweigh the technical difficulties.

\section{Conclusions}
\label{conclusions}
 In this paper we have explained that blockchain and smart contracts are independent technologies that can be used separately and congruently. We have argued that smart contracts and blockchains are not intrinsically associated or necessarily associated but that their association has been conflated by the development of cryptocurrency platforms; consequently the  drawbacks that afflict blockchains and questions about whether they are    a useful technology or only a technology bubble, do not extend 
 to smart contracts. Smart contracts are a technology in their own right and with unquestionable potential to handle much more complex scenarios that the hypothetical use case outlined in the paper.
  
We have pointed out that smart contract applications can be 
implemented with smart contracts deployed on--chain, 
on--TTP, and even, on architectures that combine the use of both
on--chain and on--TTP smart contracts. We have
argued that in spite of all the inconveniences that
TTPs introduce (single point of failures, centralised trust
and so on) on--TTP smart contracts are not necessarily 
a bad idea to be ruled out.
TTPs bring features (simplicity, centralised authority,
performance and so on) that can prove to be valuable in
the implementation of pragmatic solutions, naturally, only
in applications that can tolerate the inconveniences that
TTPs inevitably introduce. 
We have used the car insurance application shown 
in Fig.~\ref{fig:carinsuranceapplication} to present
our arguments. We use it to illustrate
the technical difficulties that a developer would face
if he or she opted for an on--chain approach. 
 It is sensible to expect that different applications will pose 
 different requirements and difficulties, yet we believe 
 that some of the issues that we faced in this scenario 
 will replicate in others.

Table~\ref{tab:summary} 
summarises our analysis. On the basis of this  analysis we have 
concluded that  an  on--TTP  implementation  of the example is 
significantly simpler that its an on--chain equivalent. Therefore, 
our piece of advice for developers is the following:
 
\begin{itemize}
\item Not to overrate the decentralised features of blockchains;
  build on their grounds only when the benefits that they bring
  clearly outweigh the technical complexities.
\item Not to underrate the conveniences that TTPs bring when 
  it comes to implement practical applications. 
 \item  We acknowledge that some application cannot
 be implemented naturally by on--chain or on--TTP smart contracts
 in isolation. Our advice in this cases is to consider the 
 use of hybrid architectures that intent to combine the advantages of 
  both approaches.
 \item To be cognisant of the necessary dispute resolution 
  mechanisms and access to justice routes that any solutions must 
  provide parties with to seek remedies.
\end{itemize}
 
\begin{table}[]
  \begin{center}

    \begin{tabular}{|c|c|c|}
      \hline
      \textbf{Issue} & \textbf{on--chain} & \textbf{on--TTP} \\
      \hline
      \hline
       Encryption    & \makecell{Hard to manage.}      
                     & \makecell{Manageable.}  \\
      \hline
       GDPR compliance    & \makecell{Extremely hard.}      
                          & \makecell{Manageable.}  \\
       \hline
       Gas cost       & \makecell{Might become \\
                                  unaffordable.} 
                      & \makecell{TTP charges \\
                                  might apply.}  \\
       \hline
       Block size     & \makecell{Might manifest.}  
                      & \makecell{Depends on the \\ TTP.}  \\
       \hline
        Direct API calls & \makecell{Not supported, \\ mediators
                                     like oracles \\ needed.} 
                         & \makecell{Supported.}  \\
       \hline
       Data inconsistencies & \makecell{Exception handling \\ 
                                mechanisms needed.}  
                            & \makecell{No risk.}  \\
       \hline
  \end{tabular}
   \vspace{0.25 cm}
   \caption{Advantages and disadvantages of on--chain and
             on--TTP implementations.}
   \label{tab:summary}
  \end{center}
\end{table}

\section*{Acknowledgements} 
Ioannis Sfykaris was partly supported by the EU Horizon 2020 project 
PrismaCloud (https://prismacloud.eu) under GA No. 644962.

\bibliographystyle{IEEEtran}
\bibliography{./biblio/bibliography.bib}

\begin{thebibliography}{1}\setlength{\itemsep}{-1ex}\small

\bibitem{ex1}
I.~M. Author.
\newblock Some related article {I} wrote.
\newblock {\em Some Fine Journal}, 99(7):1--100, January 1999.

\bibitem{ex2}
A.~N. Expert.
\newblock {\em A Book He Wrote}.
\newblock His Publisher, Erewhon, NC, 1999.

\end{thebibliography}


\begin{thebibliography}{10}
\providecommand{\url}[1]{#1}
\csname url@samestyle\endcsname
\providecommand{\newblock}{\relax}
\providecommand{\bibinfo}[2]{#2}
\providecommand{\BIBentrySTDinterwordspacing}{\spaceskip=0pt\relax}
\providecommand{\BIBentryALTinterwordstretchfactor}{4}
\providecommand{\BIBentryALTinterwordspacing}{\spaceskip=\fontdimen2\font plus
\BIBentryALTinterwordstretchfactor\fontdimen3\font minus
  \fontdimen4\font\relax}
\providecommand{\BIBforeignlanguage}[2]{{%
\expandafter\ifx\csname l@#1\endcsname\relax
\typeout{** WARNING: IEEEtran.bst: No hyphenation pattern has been}%
\typeout{** loaded for the language `#1'. Using the pattern for}%
\typeout{** the default language instead.}%
\else
\language=\csname l@#1\endcsname
\fi
#2}}
\providecommand{\BIBdecl}{\relax}
\BIBdecl

\bibitem{Satoshi2008}
S.~Nakamoto, ``Bitcoin: A peer-to-peer electronic cash system,'' \url
  {http://nakamotoinstitute.org/bitcoin/}, Visited 13 Nov 2017 2008.

\bibitem{Hileman2017}
G.~Hileman and M.~Rauchs, ``The global cryptocurrency benchmarking study,''
  \url
  {www.jbs.cam.ac.uk/fileadmin/user_upload/research/centres/alternative-finance/downloads/2017-global-cryptocurrency-benchmarking-study.pdf},
  Visited 13 Nov 2017 2017.

\bibitem{Ethereum2017}
Ethereum, ``A next-generation smart contract and decentralized application
  platform,'' \url {https://github.com/ethereum/wiki/wiki/White-Paper}, Visited
  23 Oct 2017 2017.

\bibitem{HyperledgerHome}
{The Linux Foundation}, ``Hyperledger,'' \url {www.hyperledger.org}, Visited
  Nov 2017 2017.

\bibitem{BitcoinHome}
{Bitcoin Project}, ``Bitcoin: Bitcoin is an innovative payment network and a
  new kind of money,'' \url {https://bitcoin.org}, 2019.

\bibitem{Morgen2018}
M.~E. Peck, ``Do you need a blockchain?: This chart will tell you if the
  technology can solve your problem,'' \emph{IEEE Spectrum}, vol.~54, no.~3,
  pp. 38--60, Aug 2018.

\bibitem{Scriber2018}
B.~A. Scriber, ``A framework for determining blockchain applicability,''
  \emph{IEEE Software}, vol.~35, no.~4, pp. 70--77, Jul/Aug 2018.

\bibitem{Valentina2018}
V.~Gatteschi, F.~Lamberti, C.~Demartini, C.~Pranteda, and V.~Santamaria, ``To
  blockchain or not to blockchain: That is the question,'' \emph{IP
  Professional}, vol.~20, no.~2, pp. 62--74, Mar/Apr 2018.

\bibitem{KenjiSaito2016}
K.~Saito and H.~Yamada, ``What’s so different about blockchain?
  –-blockchain is a probabilistic state machine-–,'' in \emph{Proc. IEEE
  36th Int'l Conf. on Distributed Computing Systems Workshops}, 2016.

\bibitem{Szabo1997}
N.~Szabo, ``Smart contracts: Formalizing and securing relationships on public
  networks,'' \emph{First Monday}, vol.~2, no.~9, Sep. 1997.

\bibitem{AndreasAntonopoulos2017}
A.~Antonopoulos, \emph{Mastering Bitcoin}, 2nd~ed.\hskip 1em plus 0.5em minus
  0.4em\relax O'Reilly, 2017.

\bibitem{BitcoinScript}
{Bitcoin Wiki}, ``Scrypt,'' \url {https://en.bitcoin.it/wiki/Script}, 2018.

\bibitem{AndreasAntonopoulos2018}
A.~M. Antonopoulos and D.~G. Wood, \emph{Mastering Ethereum}.\hskip 1em plus
  0.5em minus 0.4em\relax O'Reilly, 2019.

\bibitem{Solidity2017}
Ethereum, ``Solidity,'' \url
  {http://solidity.readthedocs.io/en/develop/index.html}, Visited 23 Oct 2017
  2017.

\bibitem{Minsky1985}
N.~H. Minsky and A.~D. Lockman, ``Ensuring integrity by adding obligations to
  privileges,'' in \emph{Proc. 8th Int'l Conf. on Software Engineering}, 1985,
  pp. 92--102.

\bibitem{Lindsay1993}
L.~F. Marshall, ``Representing management policy using contract objects,'' in
  \emph{Proc. IEEE First Int'l Workshop on Systems Management}, 1993, pp.
  27--30.

\bibitem{Mainelli2015}
M.~Mainelli, ``In third parties we (mis)trust?'' \emph{FinTech Futures}, 17th
  Dec 2015.

\bibitem{VictorCostan2016}
V.~Costan and S.~Devadas, ``Intel sgx explained,'' \url
  {https://eprint.iacr.org/2016/086.pdf}, 2016.

\bibitem{Sandro2019}
S.~Pinto and N.~Santos, ``Demystifying arm trustzone: A comprehensive survey,''
  \emph{ACM Computing Surveys}, vol.~51, no. 6, article 130, Jan. 2019.

\bibitem{Gorbunov2018}
S.~Gorbunov and S.~Micali, ``Pure proof-of-stake blockchains: Secure blockchain
  decentralization via committees,'' \url
  {https://medium.com/algorand/secure-blockchain-decentralization-via-committees-7602f598a0a9},
  Mar 2018, visited on 30 Mar 2019.

\bibitem{carInsuranceExample2019}
L.~Song, ``{SoliditySample},'' \url
  {https://github.com/songlinm/SoliditySample}, Visited on 18 Feb 2019,
  deployed on Ethereum blockchain at
  0xb3C66fA11af5b4975D74C654665A0b7E505b2bDe.

\bibitem{metamask}
{metamask support}, ``metamask,'' \url {https://metamask.io}, Visited 24 Jul
  2018 2018.

\bibitem{truffleHome}
Consensys, ``Truffle: Sweet tools for smart contracts,'' \url
  {https://truffleframework.com}, Visited 21 Dec 2018 2018.

\bibitem{RustHomepage2018}
``Rust,'' \url {https://www.rust-lang.org}, 2018.

\bibitem{GolangHomepage2018}
``Go: Documentation,'' \url {https://golang.org/doc/}, 2018.

\bibitem{GethEth}
Ethereum, ``Geth \& eth: Command line tools for the ethereum network,'' \url
  {https://ethereum.org/cli}, Visited 29 Jun 2018 2018.

\bibitem{infuraHome}
{CONSENSYS, AG}, ``Infura: Your access to the ethereum network,'' \url
  {https://infura.io}, Visited 13 Nov 2018 2018.

\bibitem{carInsuranceExampleEtherscan2019}
L.~Song, ``{CarInsurancePolicyManager deployed on Ethereum},'' \url
  {https://etherscan.io/address/0xb3c66fa11af5b4975d74c654665a0b7e505b2bde},
  Visited on 18 Feb 2019.

\bibitem{GDPR2018}
{The european parliament and the council of the european union}, ``General data
  protection regulation,'' \url
  {https://eur-lex.europa.eu/legal-content/EN/TXT/PDF/?uri=CELEX:32016R0679&from=EN},
  Apr 2016.

\bibitem{GDPRinfo2018}
{Intersoft Consulting}, ``General data protection regulation: Gdpr,'' \url
  {https://gdpr-info.eu}, 2018.

\bibitem{JuanBenet2014}
``Ipfs - content addressed, versioned, p2p file system (draft 3),'' \url
  {https://arxiv.org/pdf/1407.3561.pdf}, Jul 2014, arXiv:1407.3561 [cs.NI].

\bibitem{ipfs}
{IPFS}, ``Ipfs is the distributed web,'' \url {https://github.com/ipfs/ipfs},
  2019, visited on 16 Feb 2019.

\bibitem{EthereumYellowPaper2019}
D.~Wood, ``Ethereum: A secure decentralised generalised transaction ledger
  byzantium version 4e05aa0 - 2019-03-04,'' \url
  {https://github.com/ethereum/yellowpaper}, 2019, yellow Paper. Visited on 12
  Mar 2019.

\bibitem{EthreumGasLimitChart2018}
Etherscan, ``Chart ethereum gaslimit history,'' \url
  {https://etherscan.io/chart/gaslimit}, 2018, visited on 28 Feb 2018.

\bibitem{Cryptostats2018}
Bitinfocharts, ``Cryptocurrency statistics,'' \url {https://bitinfocharts.com},
  2018, visited on 28 Feb 2018.

\bibitem{Andrew2017}
A.~Miller, I.~Bentov, R.~Kumaresan, C.~Cordi, and P.~McCorry, ``Sprites and
  state channels: Payment networks that go faster than lightning,'' \url
  {https://arxiv.org/pdf/1702.05812.pdf}, Nov 2017, arXiv:1702.05812 [cs.CR].

\bibitem{Gideon2016}
G.~Greenspan, ``Why many smart contract use cases are simply impossible,'' \url
  {https://www.coindesk.com/three-smart-contract-misconceptions/}, 16 Apr 2016,
  visited on 9 Nov 2018.

\bibitem{Fan2016}
F.~Zhang, E.~Cecchetti, K.~Croman, A.~Juels, and E.~Shi, ``Town crier: An
  authenticated data feed for smart contracts,'' in \emph{Proc. 23rd ACM Conf.
  on Computer and Communications Security (CCS’16)}, 2016.

\bibitem{Davidhamilton2018}
D.~Hamilton, ``What are bitcoin oracles? new functionality on the blockchain,''
  \url {https://coincentral.com/what-are-bitcoin-oracles/}, 30 Aug 2018,
  visited on 9 Nov 2018.

\bibitem{oraclizeHome}
{Oraclize Limited}, ``Oraclize,'' http://www.oraclize.it, 2018.

\bibitem{Decker2016}
C.~Decker, J.~Seidel, and R.~Wattenhofer, ``Bitcoin meets strong consistency,''
  in \emph{Proc. 17th Int'l Conf. on Distributed Computing and Networking
  (ICDCN'16)}, 2016.

\bibitem{GavinWood2015}
G.~Wood, ``Chain reorganisation depth expectations,'' \url
  {https://blog.ethereum.org/2015/08/08/chain-reorganisation-depth-expectations/},
  Aug 2015, visited on 18 Feb 2019.

\bibitem{Trent2016}
T.~McConaghy, R.~Marques, A.~M{\"{u}}ller, D.~D. Jonghe, T.~T. McConaghy,
  G.~McMullen, R.~Henderson, S.~Bellemare, and A.~Granzotto, ``Bigchaindb: A
  scalable blockchain database,'' \url
  {www.bigchaindb.com/whitepaper/bigchaindb-whitepaper.pdf}, Visited 1 Nov 2017
  2017.

\bibitem{MolinaTSC2011}
C.~Molina-Jimenez, S.~Shrivastava, and M.~Strano, ``A model for checking
  contractual compliance of business interactions,'' \emph{IEEE Trans. on
  Service Computing}, vol.~PP, no.~99, 2011.

\bibitem{Solaiman2015}
E.~Solaiman, I.~Sfyrakis, and C.~Molina-Jimenez, ``Dynamic testing and
  deployment of a contract monitoring service,'' in \emph{Proc. 5th Int'l Conf.
  on Cloud Computing and Services Science (CLOSER'15)}, 2015.

\bibitem{conch}
C.~Molina-Jimenez, ``{conch},'' \url {https://github.com/carlos-molina/conch},
  Visited in Oct 2018.

\bibitem{Drools5.3.0UserGuide}
{The JBoss Drools team}, ``Drools expert user guide,'' \url
  {https://docs.jboss.org/drools/release/5.3.0.Final/drools-expert-docs/html/index.html},
  visited: 7 May 2018.

\bibitem{Myong2001}
M.~H. Kang, J.~S. Park, and J.~N. Froscher, ``Access control mechanisms for
  inter-organizational workflow,'' in \emph{Proc. Sixth ACM symposium on Access
  control models and technologies (SACMAT'01)}, 2001.

\bibitem{Bethencourt2007}
J.~Bethencourt, A.~Sahai, and B.~Waters, ``Ciphertext-policy attribute--based
  encryption,'' in \emph{Proc. IEEE Symposium on Security and Privacy (SP'07)},
  2007.

\bibitem{Viveki2003}
V.~Vishnumurthy, S.~Chandrakumar, and E.~G. Sirer, ``Karma: A secure economic
  framework for peer-to-peer resource sharing,'' in \emph{Proc. 1ST Workshop on
  economics of peer-to-peer systems}, 2003.

\bibitem{Douceur2002}
J.~R. Douceur, ``The sybil attack,'' in \emph{Proc. First Int'l Workshop,
  IPTPS(2002)}, 2002, pp. 251--260.

\bibitem{codelegit2019}
{Codelegit}, ``Technical compliance,'' \url {http://codelegit.com}, 2019,
  visited on 9 Feb 2019.

\bibitem{Stefan2019}
S.~Dziembowski, L.~Eckey, S.~Faust, and D.~Malinowski, ``Perun: Virtual payment
  hubs over cryptographic currencies,'' in \emph{Proc. 40th IEEE Symposium on
  Security and Privacy (IEEE S and P)}, 2019.

\bibitem{PayPalDisputeResolution}
{Pay Pal}, ``Resolving a dispute with your seller guide to handling a
  dispute,'' \url {https://www.paypal.com/uk/webapps/mpp/first-dispute}, 2019,
  visited on 9 Feb 2019.

\bibitem{AmazonPayDisputeResolution}
{Amazon Pay}, ``Transaction disputes,'' \url
  {https://pay.amazon.com/uk/help/201754740}, 2019, visited on 9 Feb 2019.

\bibitem{Rongbing2015}
R.~Liu, ``The role of alipay in china,'' Master's thesis, Information Science,
  Master programme, Radboud University Nijmegen, 2015.

\bibitem{Clement2018}
C.~Lesaege and F.~Ast, ``Kleros short paper (v1.0.6),'' \url
  {https://kleros.io/assets/whitepaper.pdf}, Nov 2018, visited on 9 Feb 2019.

\bibitem{Mateja2018}
M.~Durovic, ``Law and autonomous systems series: How to resolve smart contract
  disputes - smart arbitration as a solution,'' \url
  {https://www.law.ox.ac.uk/business-law-blog/blog/2018/06/law-and-autonomous-systems-series-how-resolve-smart-contract-disputes},
  Jun 2018.

\bibitem{Kieron2017}
K.~O'Hara, ``Smart contracts-- dumb idea,'' \emph{IEEE Internet Computing},
  vol.~21, no.~2, 2017.

\bibitem{Jimmy2018}
J.~Song, ``The truth about smart contracts,'' \url
  {https://medium.com/@jimmysong/the-truth-about-smart-contracts-ae825271811f},
  11 Jun 2018, visited on 9 Nov 2018.

\bibitem{Cubas2018}
R.~C. Celiz, Y.~E. D.~L. Cruz, and D.~M. Sanchez, ``Cloud model for purchase
  management in health sector of peru based on iot and blockchain,'' in
  \emph{Proc. IEEE 9th Annual Information Technology, Electronics and Mobile
  Communication Conf. (IEMCON)}, 2018.

\bibitem{MolinaSOCA2011}
C.~Molina-Jimenez, S.~Shrivastava, and S.~Wheater, ``An architecture for
  negotiation and enforcement of resource usage policies,'' in \emph{Proc. IEEE
  Int'l Conf. on Service Oriented Computing \& Applications (SOCA 2011)}, 2011.

\bibitem{Minsky1991}
N.~H. Minsky, ``Law-governed systems,'' \emph{Software Engineering Journal},
  vol.~6, no.~5, 1991.

\bibitem{MinskySerban2005}
N.~Minsky, ``Law governed interaction (lgi): A distributed coordination and
  control mechanism (an introduction, and a reference manual),'' \url
  {http://www.moses.rutgers.edu/documentation/manual.pdf}, 2005, october 24,
  2005 Version 0.9.2.

\bibitem{ludwig2003soa}
H.~Ludwig and M.~Stolze, ``Simple obligation and right model {(SORM)}-for the
  runtime management of electronic service contracts,'' in \emph{Proc. 2nd
  Int'l Workshop on Web Services, e--Business, and the Semantic Web(WES'03),
  LNCS vol. 3095}, 2003, pp. 62--76.

\bibitem{PedroGama2006}
P.~Gama, C.~Ribeiro, and P.~Ferreira, ``Heimdhal: A history--based policy
  engine for grids,'' in \emph{Proc. 6th IEEE Int'l Symp. on Cluster Computing
  and the Grid (CCGRID'06)}.\hskip 1em plus 0.5em minus 0.4em\relax IEEE CS,
  2006, pp. 481--488.

\bibitem{perringodart}
O.~Perrin and C.~Godart, ``{An approach to implement contracts as trusted
  intermediaries},'' in \emph{Proc. 1st IEEE Int'l Workshop on Electronic
  Contracting (WEC'04)}, 2004, pp. 71--78.

\bibitem{Marko2015}
M.~Vukoli{\'{c}}, ``The quest for scalable blockchain fabric: Proof-of-work vs.
  bft replication,'' in \emph{Proc. Int'l Workshop on Open Problems in Network
  Security (iNetSec'15), LNCS}, 2015, pp. 112--125.

\bibitem{Shebar2017}
S.~Bano, A.~Sonnimo, M.~Al-Bassam, S.~Azouvi, P.~McCorry, S.~Meiklejohn, and
  G.~Danezis, ``Sok: Consensus in the age of blockchains,'' \url
  {https://arxiv.org/pdf/1711.03936.pdf}, Nov 2017, arXiv:1711.03936v2.

\bibitem{Joseph2016}
J.~Poon and T.~Dryja, ``The bitcoin lightning network: Scalable off-chain
  instant payments,'' \url
  {https://lightning.network/lightning-network-paper.pdf}, Jan. 2016.

\bibitem{CarlosIoannisTurin2018}
C.~Molina-Jimenez, E.~Solaiman, I.~Sfyrakis, I.~Ng, and J.~Crowcroft, ``On and
  off-blockchain enforcement of smart contracts,'' in \emph{Proc. Int'l
  Workshop on Future Perspective of Decentralized Applications (FPDAPP), LNCS
  Vol. 11339}, 2018, pp. 342--354.

\bibitem{Eberhardt2017}
J.~Eberhardt and S.~Tai, ``On or off the blockchain? insights on off-chaining
  computation and data,'' in \emph{(ESOCC'17)}, 2017.

\bibitem{CarlosIoannisSC22018}
C.~Molina-Jimenez, I.~Sfyrakis, E.~Solaiman, I.~Ng, A.~Chun, and J.~Crowcroft,
  ``Implementation of smart contracts using on-- and off--blockchain
  components,'' in \emph{Proc. 8th IEEE Int'l Symposium on Cloud and Services
  Computing (SC2)}, 2018.

\bibitem{GuyZyskind_enigma2015}
G.~Zyskind, O.~Nathan, and A.~S. Pentland, ``Enigma: Decentralized computation
  platform with guaranteed privacy,'' \url {https://arxiv.org/abs/1506.03471}
  (visited in Mar 2018), Jan 2015, arXiv:1506.03471v1 [cs.CR].

\bibitem{Florian2016}
F.~Idelberger, G.~Governatori, R.~Riveret, and G.~Sartor, ``Evaluation of
  logic--based smart contracts for blockchain systems,'' in \emph{Proc. 10th
  Int'l Symposium RuleML'16: Rule Technologies: Research, Tools, and
  Applications, LNCS, Vol 9718}, 2018, pp. 167–183,.

\bibitem{Zilliqa2017}
ZILLIQA, ``The {ZILLIQA} technical whitepaper,'' \url
  {https://docs.zilliqa.com/whitepaper.pdf}, ZILLIQA, Tech. Rep. Version 0.1,
  Aug 2017.

\end{thebibliography}

\end{document}